\documentclass[]{raa}            
\usepackage{graphicx,times}
\usepackage{natbib}
\providecommand{\bm}[1]{\mbox{\boldmath$#1$\unboldmath}}

\begin{document}
\title{Physical origin of multi-wavelength emission of GRB 100418A and implications for its progenitor
\footnotetext{\small Supported by the National Natural Science Foundation of China.}
}

 \volnopage{ {\bf 0000} Vol.\ {\bf 0} No. {\bf XX}, 000--000}
   \setcounter{page}{1}

   \author{Lan-Wei Jia\inst{1,2}
       \and  Xue-Feng Wu\inst{3}
   \and Hou-Jun L\"{u}\inst{4,2}
   \and Shu-Jin Hou \inst{5,2}
    \and En-Wei Liang\inst{*2}
   }

   \institute{Department of Physics, Guangzhou University, Guangzhou 510006, China\\
        \and
             Department of Physics and GXU-NAOC Center for Astrophysics
and Space Sciences, Guangxi University, Nanning 530004,
China; lew@gxu.edu.cn \\
       \and
       Purple Mountain Observatory, Chinese Academy of Sciences, Nanjing 210008, China\\
       \and
       Department of Physics and Astronomy, University of Nevada Las Vegas, Las Vegas, NV 89154, USA\\
       \and
       Department of Physics and Institute of Theoretical Physics and Astrophysics, Xiamen University, Xiamen, 361005, China\\
\vs \no
   {\small Received [2011] [March] [4]; accepted [2011] [December] [20] }
}

\abstract{ {\it Swift} GRB 100418A is a long burst at $z= 0.624$ without detection of any associated supernova (SN). Its lightcurves in both the prompt and afterglow phases are similar to GRB 060614, a nearby long GRB without an associated SN. We analyze the observational data of this event and discuss the possible origins of its multi-wavelength emission. We show that its joint lightcurve at 1 keV derived from {\it Swift} BAT and XRT observations is composed of two distinguished components. The first component, whose spectrum is extremely soft ($\Gamma = 4.32$), ends with a steep decay segment, indicating the internal origin of this component. The second component is a slowly-rising, broad bump which peaks at $\sim 10^5$ seconds post the BAT trigger. Assuming that the late bump is due to onset of the afterglow, we derive the initial Lorentz factor ($\Gamma_0$) of the GRB fireball and find that it significantly deviates from the relation between the $\Gamma_0$ and isotropic gamma-ray energy derived from typical GRBs. We also check whether it follows the same anti-correlation between X-ray luminosity and the break time observed in the shallow decay phase of many typical GRBs, which is usually regarded as a signal of late energy injection from the GRB central engine. However, we find that it does not obey this correlation. We propose that the late bump could be contributed by a two-component jet. We fit the second component with an off-axis jet model for a constant medium density and find the late bump can be represented by the model. The derived jet half-opening angle is $0.30$ rad and the viewing angle is $0.315$ rad. The medium density is $0.05$ cm$^{-3}$, possibly suggesting that it may be from a merger of compact stars. The similarity between GRBs 060614 and 100418A may indicate that the two GRBs are from the same population and the late bump observed in the two GRBs may be a signal of a two-component jet powered by the GRB central engine.
}

   \authorrunning{L. W. Jia et al. }            
   \titlerunning{Multi-wavelength Emission of GRB 100418A and implications}  
   \maketitle


%
%
\section{Introduction}           
\label{sect:intro}
Phenomenologically, two types of gamma-ray bursts (GRBs) have been classified according to their
observed burst duration $T_{90}$ based on observations with CGRO/BATSE (Kouveliotou et al. 1993).
Recent progress made by the {\it Swift} mission presents several lines of evidence that burst duration is no longer a reliable indicator for GRB classification (Zhang 2006; Zhang et al. 2007, 2009; L\"{u} et al. 2010; Xin et al. 2011). The most prominent case is GRB 060614, which is a long GRB at redshift $z=0.125$. It is surprising that no supernova is
associated with this nearby long GRB (Gal-Yam et al. 2006; Gehrels et al. 2006;
Della Valle et al. 2006; Fynbo et al. 2006) as seen in other nearby long GRBs 980425, 030329,
031203, and 060218 (Galama et al. 1998; Hjorth et al. 2003; Stanek et al. 2003;
Modjaz et al. 2006; Pian et al. 2006), leading to debate on the physical origin
of this event, i.e., collapse of a massive star (Type II) or merger of a compact
star binary (Type I) (e.g., Zhang 2006 and reference therein). Some intrinsically
short-duration, high-$z$ GRBs, such as GRBs 080913 ($z = 6.7$; Greiner et al.
2009) and 090423 ($z = 8.3$; Tanvir et al. 2009; Salvaterra et al. 2009), and a
typical short-duration high-$z$ GRB 090426 (Levesque et al. 2010; Xin et al.
2011) suggest that some short duration GRBs are probably not produced via
compact star mergers (Type I), but are likely related to massive stars (Type
II). Virgili et al. (2011) performed a series of Monte Carlo simulations and
showed that the compact star merger model cannot interpret both the {\it Swift}
known-$z$ short GRB sample and the {\it CGRO}/BATSE short GRB sample. Zhang et
al. (2009) attempted to invoke a set of multiple observational criteria to judge
the physical category of a GRB, and gave an operational procedure to discern
the physical origin of GRBs. Based on observed gamma-ray energy ($E_{\rm
iso}$) and peak energy ($E_{\rm p}$) of the $\nu F_{\nu}$ spectrum of prompt
gamma-ray emission, L\"{u} et al. (2010) defined a parameter $\varepsilon
\equiv E_{\rm iso}/E_{\rm p,z,}^{1.7}$, and proposed a new empirical classification
method that is found to better match the physically-motivated Type II/I
classification scheme. They showed that the typical Type II GRBs are in the
high-$\varepsilon$ group, in contrast to the typical type I GRBs, which belong
to the low-$\varepsilon$ group. The non-detection of any SN associated with GRB
060614 also motivated ideas that it may have an essentially different
physical origin from both the Type I and II, such as a stellar object disrupted
by a median-mass black hole (Lu et al. 2008). 060614-like GRBs are of
interest to study the physical origin of these kinds of events.

GRB 100418A interestingly triggered Swift/BAT. It is quite similar to GRB 060614, possibly adding a valuable case for such kinds of events. This paper presents a detailed analysis of this event and compares it with GRB 060614, hence discussing
possible physical origins of the multi-wavelength emission of this event. Throughout, a concordance cosmology
with parameters $H_0 = 71$ km s$^{-1}$ Mpc$^{-1}$, $\Omega_M=0.30$, and
$\Omega_{\Lambda}=0.70$ is adopted.

\section{Data}

\subsection{Prompt $\gamma$-Rays}
GRB 100418A triggered the {\it Swift} satellite on 2010 April 18 at $T_0=$21:10:08
UT. The BAT light curve shows two overlapping peaks starting at $T_0-10$ sec,
peaking at $T_0+2$ sec, and ending at $T_0+40$ sec. Its $T_{90}$ is 7$\pm$1 sec in
the 15-350 keV band, with weak extended emission up to roughly 40 seconds post
the BAT trigger (Ukwatta et al. 2010). The time-averaged spectrum from $T_0-1.1$ sec
to $T_0+7.8$ sec is best fit with a simple power-law model, yielding a power law photon
index of $\Gamma=2.16\pm 0.25$. Although its $E_{\rm p}$ cannot be measured
directly from the observed BAT spectrum, a photon index of 2.16 is very close
to the typical high energy photon index (Kaneko et al. 2006), indicating that
$E_{\rm p}$ would be lower than 50 keV. Using the $\Gamma$ - $E_{\rm p}$ relation (Zhang
et al. 2007; Sakamoto et al. 2009) and Bayesian methodology (Butler \&
Kocevski. 2007) one may obtain $E_{\rm p} = 29^{+2}_{-27}$ keV. The fluence in the
15-150 keV band is $(3.4\pm0.5)\times10^{-7}$ erg cm$^{-2}$. The 1-sec peak
photon flux measured from $T_0+0.47$ sec in the 15-150 keV band is 1.0$\pm$0.2 ph
cm$^{-2}$ s$^{-1}$. Assuming that the low and high energy band spectral indices
are $-1$ and $-2.3$, one can get $E_{\rm iso}$ and the peak luminosity
($L_{\gamma,\rm iso})$ in the $1 - 10^4$ keV band are $ 9.9^{+6.3}_{-3.4}
\times 10^{50}$ erg and $2.1^{+1.1}_{-0.6} \times 10^{50}$ erg
s$^{-1}$, respectively, with redshift $z = 0.624$ (Marshall et al. 2011).

\subsection{Afterglows}
The XRT began observing the field of GRB 100418A at 79.1 s after the BAT
trigger. We use the web-based analysis system at http://www.swift.ac.uk/ for XRT data analysis. Details of the system are available in Evans et al. (2007). It is found that the XRT lightcurve starts with a steep decay segment, then transits to a smooth bump peaking at $\sim 10^5$ seconds post the GRB trigger. The spectrum of the early steep decay phase is extremely soft. It can be fit with an absorbed power-law. The derived photon index is $\Gamma =4.32^{+0.28}_{-0.24}$ and the intrinsic column density of the host galaxy is $N_{H}=(2.1\pm0.4) \times$ $10^{21}$ cm$^{-2}$ over the Galactic absorption ($4.8\times10^{20}$ cm$^{-2}$). The C-stat of the fit is 206 in 206 degrees of freedom. The late X-ray spectrum accumulated in the period of $\sim 10^3 - 10^5$ seconds post the GRB trigger, however, is similar to that observed in typical GRBs (Liang et al. 2007), which can be fitted with an absorbed power-law with a photon index of $\Gamma =2.04^{+0.21}_{-0.29}$ and an intrinsic absorption of $N_{H}=(1.79^{+0.93}_{-1.51}) \times$ $10^{21}$ cm$^{-2}$. The C-stat of the fit is 119 in 131 degrees of freedom. No significant difference of $N_H$ is found in the early and late epochs.

The optical afterglow was detected in the white, b, u and v filters (Siegel \&
Marshall. 2010). Optical data are collected from GCN circulars. They are
not corrected for the Galactic extinction corresponding to a reddening of
$E_{B-V} = 0.07$.

\section{Joint Temporal Analysis of Prompt and Afterglow Emission}
In order to present a global view of the multi-wavelength observations for GRB
100418A, we show its joint lightcurve in X-ray (at 1 keV) derived from BAT and XRT
observations and the optical lightcurve in the R band in Figure 1(a). The X-ray
lightcurve is composed of two distinguished components, but the optical
emission was detected for the second component only. The optical lightcurve
traces the X-ray one for the second component. The joint lightcurves for GRB
060614 are also shown in Figure 1(a) for comparison. It is interesting that GRB
100418A is almost a mimic of GRB 060614. Although the luminosities of the first
components of the two GRBs are comparable, the luminosity of the second
component of GRB 100418A is one order of magnitude higher in the X-ray band and
almost two orders of magnitude higher in the optical band than that of GRB 060614.

The X-ray lightcurve and the optical lightcurve are fit with a two-component
smooth broken power-law model and one-component broken power-law model,
respectively. Each component is characterized as
\begin{equation}
F = F_{0}
[(\frac{t}{t_b})^{\omega\alpha_1}+(\frac{t}{t_b})^{\omega\alpha_2}]^{-1/\omega}.
\end{equation}
We fix $\omega=3$ in our fitting. The fitting results are shown in Figure 1(b) and
Table 1. It is found that the rising phase of the second component is shallower
than that usually observed in typical GRBs (Liang et al. 2009), and the decay
segment of the first component is much steeper than that of the second one.

\section{Possible Physical Origins of Multi-Wavelength Emission}

\subsection{The prompt gamma-rays}
As shown in Figure 1(b) and Table 1, the steep decay after the peak time
($\alpha_2=4.18\pm 0.18$) and extremely soft X-ray spectrum ($\Gamma =
4.32^{+0.28}_{-0.24}$) of the first component indicate that this
component may not originate from external shocks. The early steep decay
segment observed in the XRT band could be contributed by the tail emission of
the prompt gamma-rays due to the time-delay of the photons from the high latitude of
the GRB fireball, as usually seen in some typical GRBs (Liang et al. 2006; Zhang
et al. 2007, 2009).

\subsection{The afterglows}
The second component slowly rises and peaks at $\sim 10^5$ seconds post the GRB
trigger. This component was also detected in the optical bands. The decay slope
post the peak ($\alpha_2=1.64\pm 0.12$) and spectral index ($\beta_{\rm X}=1.13$) are
consistent with the closure relation of the GRB fireball in the spectral regime
$\nu_m<\nu_{\rm X}<\nu_c$, i.e., $\alpha_2=3\beta_{\rm X}/2$, where $\nu_m$ and $\nu_c$ are
the typical frequency and the cooling frequency of synchrotron radiation respectively
(Sari et al. 1998; Zhang \& M{\'e}sz{\'a}ros 2004), generally favoring the
external origin of this component.

Some models predict a smooth bump in the afterglow lightcurves. The early bump usually observed in the optical lightcurves is believed to be due to the deceleration of the GRB fireball by the surrounding medium (Sari \& Piran 1999). In this
scenario, one can derive the initial Lorentz factor ($\Gamma_{0}$) of the GRB
fireball in the thin shell case with (Sari \& Piran 1999)
\begin{equation}\label{Gamma_0}
\Gamma_0=2[\frac{3E_{\rm iso}}{32\pi
nm_pc^5 \eta t_{\rm p,z}^3} ]^{1/8}\sim 193(n\eta)^{-1/8}\times (\frac{E_{\rm
iso,52}}{t_{\rm p,z,2}^3} )^{1/8},
\end{equation}
where $n$ is the medium density surrounding the burst (in units of cm$^{-3}$), $\eta$ is the ratio of
the $E_{\gamma, \rm iso}$ to the total kinetic energy of the GRB fireball, and
$t_{\rm p,z}=t_{\rm p}/(1+z)$ is the peak time in the cosmologically local frame. Notation $Q_n$ denotes $Q/10^n$. Liang et al. (2010) discovered a tight correlation between $\Gamma_0$ and $E_{\rm iso}$. We test if the origin of the bump in GRB 100418A is due to the deceleration of the GRB fireball. We derive its $\Gamma_0$ with Eq. 2 and obtain $\Gamma_{0}=24\pm 2$ by taking $n=1$ cm$^{-3}$ and $\eta=0.2$. The $\Gamma_0$ is much lower than typical GRBs (Liang
et al. 2010). Similar to GRB 060614, it is a significant outlier of the tight
$E_{\rm iso}-\Gamma_0$ correlation, as shown in Figure 2(a).

The second possibility is the long-lasting energy injection effect. Assuming that the
energy injection behaves as $L_{\rm in}=L_0 (t/t_{\rm b})^{-q}$, we have
$\alpha_1=(q-1)+(q+2)\beta_{\rm X}/2$ and get $q\sim 0.1$, being roughly consistent with
the energy injection from a spin-down magnetar ($q\sim 0$; Dai
\& Lu 1998; Zhang \& M{\'e}sz{\'a}ros 2001; Xu et al. 2009). A canonical XRT lightcurve is detected for most
typical GRBs, and its shallow decay segment is generally consistent with the
expectation of the energy injection models (e.g., Liang et al. 2007). Dainotti
et al. (2010) derived a relation between the break time of the shallow decay
segment and the corresponding X-ray luminosity. Recently, Xu \& Huang (2011) discovered a tight correlation  among the break time, the X-ray luminosity, and the isotropic gamma-ray energy release. We also examine if GRB 100418A is consistent with these relations. We find that it is an outlier of the $L_{\rm X}-t_{\rm b}$ relation at the 90\% confidence
level, similar to GRB 060614, as shown in Figure 2(b). The derived $L_{\rm X}$ from the relation by Xu \& Huang (2011) is $\sim 5\times 10^{44}$ erg s$^{-1}$, being smaller than the observed one by a factor of 2.

A smooth bump feature may also be explained by the line-of-sight effect
(Panaitescu \& Vestrand 2008; Guidorzi et al. 2009; Margutti et al. 2010). This
requires that the GRB jet is uniform with a sharp edge, and the line of sight
is outside the jet cone. The afterglow peak then corresponds to the epoch when
the $1/\Gamma$ beaming cone of radiation enters the line of sight, and the measured $\Gamma$ is not
the initial Lorentz factor of the ejecta, but is the Lorentz factor defined by
$(\theta_v-\theta_j) = 1/\Gamma$, where $\theta_v$ and $\theta_j$ are the
viewing angle and the jet half-opening angle, respectively. It predicts that the rising
index of the lightcurve is very steep, say, $\alpha_{r} \sim (3-4)$ (Panaitescu \&
Vestrand 2008). However, the rise can be slow if the deviation of the line of sight from the jet cone is small. The line-of sight effect alone could not explain the prompt emission, the early steep decay, or the late bump feature of GRB 100418A.

Another possibility to explain the second component may be two component jet
models (e.g., Granot et al. 2006; Racusin et al. 2008; Liang et al. 2009; Inayoshi \& Tsutsui
2010). As shown by Huang et al. (2004), a co-axial two component jet model may interpret the late optical rebrightening in GRB 030723. The inner narrow jet component produces the prompt gamma-rays and the early afterglow, and the wide hollow jet component is responsible for the late afterglow. Liu et al. (2008) argued that another off-axis jet component powered by late activity of the GRB's central engine after the main burst may contribute to the late rebrightening as that shown in 060206. Considering a jet that consists of an on-axis narrow and initially highly relativistic outflow from which the prompt emission originates and a late off-axis moderately relativistic outflow that decelerates at a significantly later time and contributes to the late afterglow (Liu et al. 2008), we fit the late X-ray and optical bump  of GRB 100418A accurately with the numerical model of Huang et al. (2000) by taking into account synchrotron-self-Compton cooling of electrons. Figure 3 shows our fit to the data with the following model parameters: isotropic kinetic energy  $E_{k,\rm iso}=10^{53}$ erg, ISM number density $n=0.05$ cm$^{-3}$, jet half-opening angle $\theta_j=0.30$ rad, viewing angle $\theta_{v}=0.315$ rad, electron energy fraction $\epsilon_e=0.15$, magnetic energy fraction $\epsilon_B=10^{-4}$, and electron energy distribution index $p=2.2$.
\begin{table}[]
\caption[]{Best-fit parameters of X-ray and optical lightcurves for GRB 100418A}
\begin{minipage}{\textwidth}
\begin{center}
\begin{tabular}{cccccccccccccccccccccc}
\hline
&$F_0$              & $t_b$ & $\alpha_1$ & $\alpha_2$ & $\chi^2$/(dof) \\
&(erg cm$^{-2}$ s$^{-1}$) & ($10^4 $s)   &            &            &                \\
\hline\noalign{\smallskip}

BAT+XRT(0-1200 s)     & $(4.19\pm 0.54)\times 10^{-8}$    & $(26\pm 5)\times 10^{-2}$          & -0.52$\pm$0.09  & 4.18$\pm$0.18    &143/89   \\
XRT($1200-10^7$ s)  & $(3.25\pm 0.68)\times 10^{-13}$   & $8.96\pm 2.24$   & -0.26$\pm$0.11  &  1.64 $\pm$0.12  &36/42    \\
R Band ($1200-10^7$ s)            & $(1.38\pm 0.06)\times 10^{-12}$   & $3.36\pm 0.20$    &-0.89$\pm$0.14  &  1.31 $\pm$0.12   &492/312  \\
\hline
\end{tabular}
\end{center}
\end{minipage}
\end{table}
\begin{figure*}
\includegraphics[angle=0,scale=0.65]{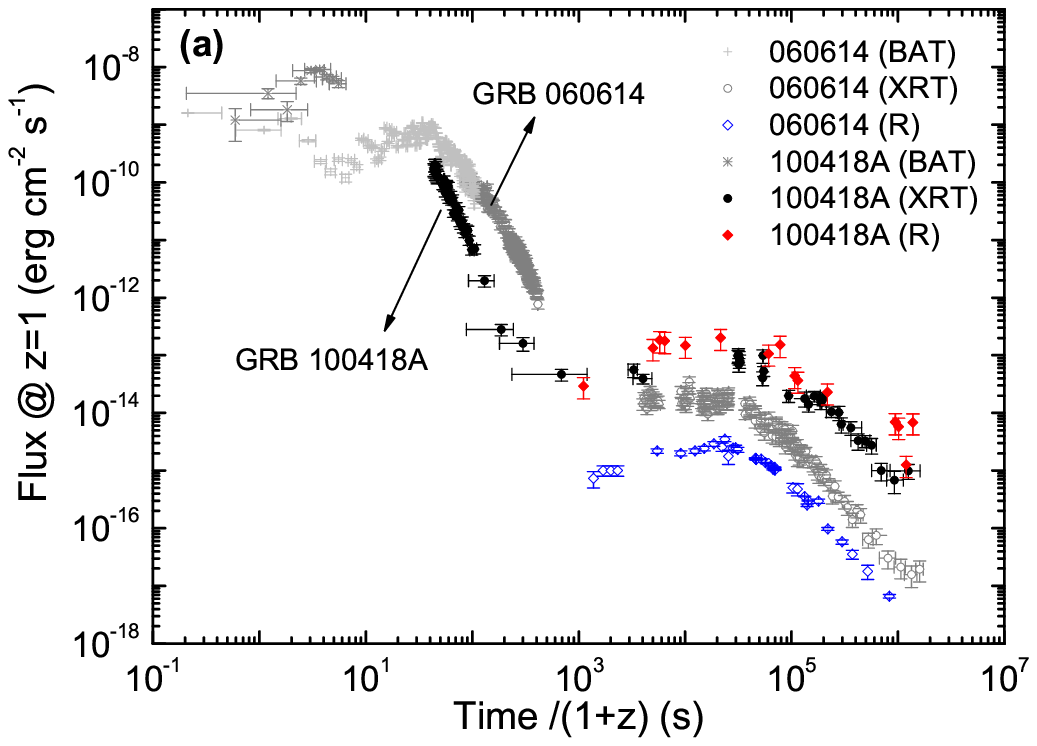}
\includegraphics[angle=0,scale=0.65]{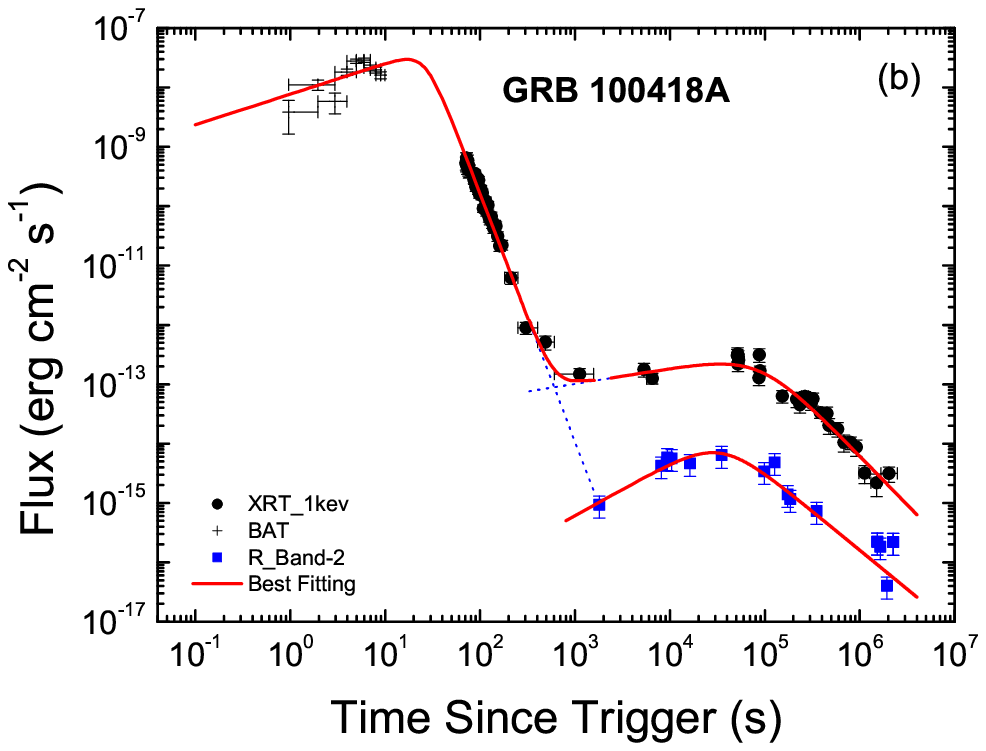}
\caption{(a) The X-ray light curve at 1 keV derived from the BAT and XRT
observations and the {\it R} band light curve of GRB 100418A in comparison with GRB
060614.(b) Best fits to the light curves of GRB 100418A with a two-component
smooth broken power-law model (lines).}
\end{figure*}

\begin{figure*}
\includegraphics[angle=0,scale=0.65]{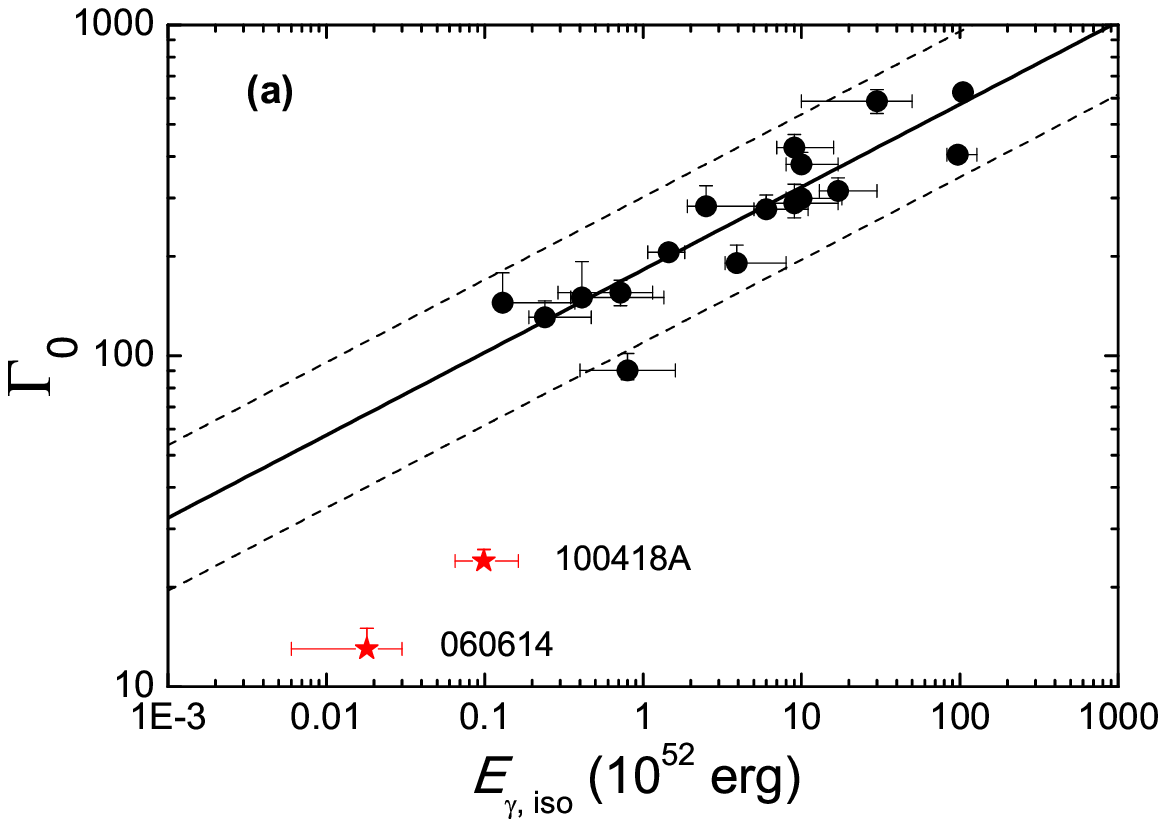}
\includegraphics[angle=0,scale=0.65]{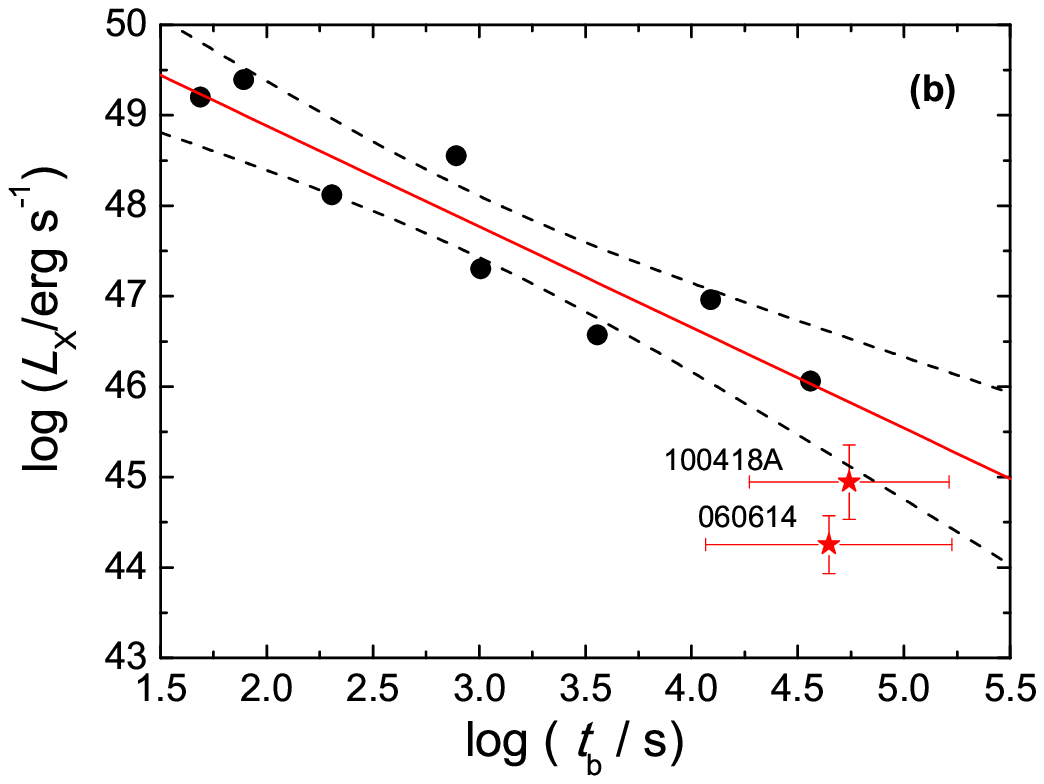}
\caption{Tests for the physical origin of the late bump of GRB 100418A with empirical relations observed in typical GRBs: (a) relation of $\Gamma_{0}$ to $E_{\rm iso}$ taken from Liang et al. (2010); (b) relation between the peak time (or break time) and the corresponding X-ray luminosity of the GRBs with a canonical XRT light curve (solid dots, from Dainotti et al. 2010). The lines are the best fit to the correlation at a confidence level of 90\%. GRBs 060614 and 100418A are marked with stars.}
\end{figure*}
\begin{figure*}
\includegraphics[angle=0,scale=0.26]{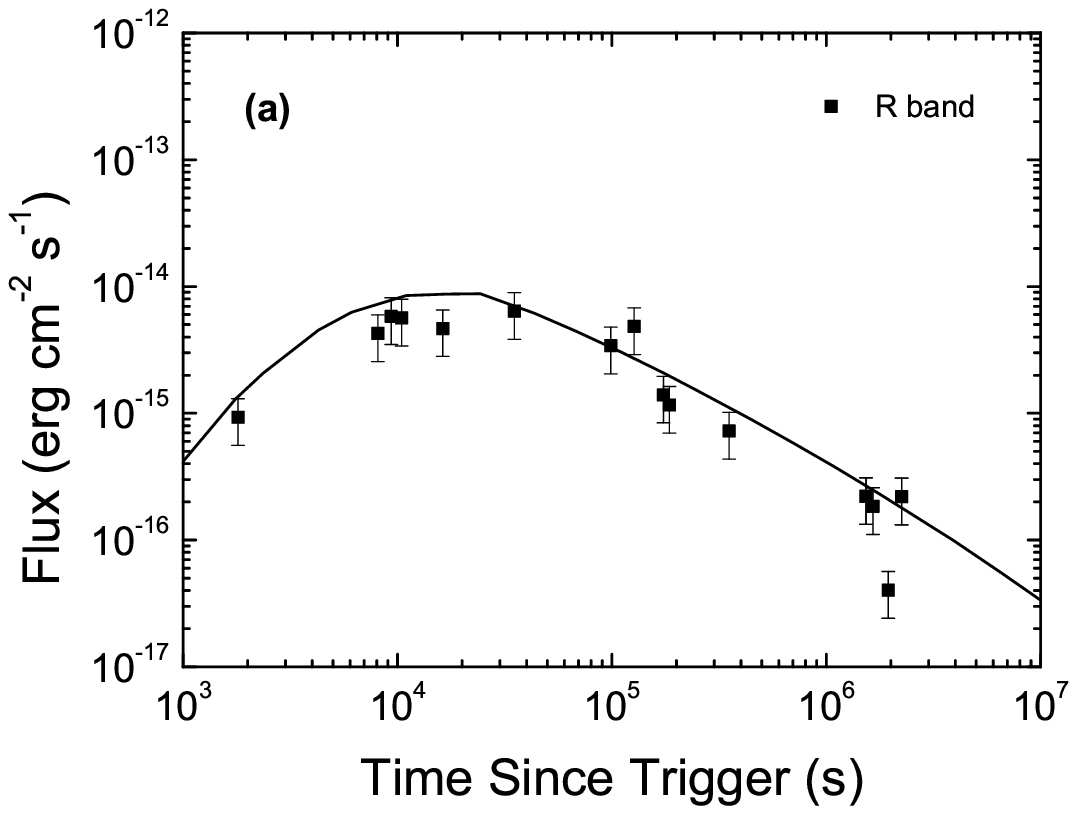}
\includegraphics[angle=0,scale=0.26]{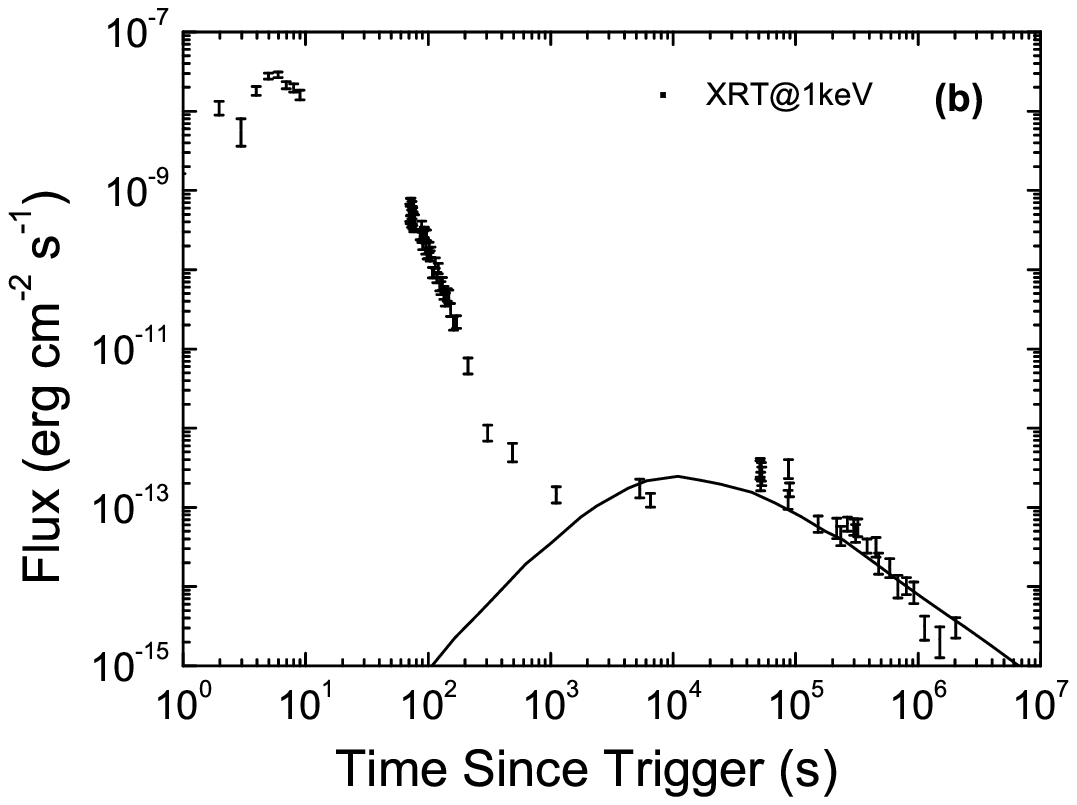}

\caption{Numerical fit to the optical (penal a) and X-ray (penal b) light curve of the second component of GRB 100418A with the afterglow model by Huang et al.(2000). The model parameters are as follows: isotropic kinetic energy $E_{k,\rm iso}=10^{53}$ erg, ISM number density $n=0.05$ cm$^{-3}$, jet half-opening angle $\theta_j=0.30$ rad, viewing angle $\theta_{v}=0.315$ rad, electron energy fraction $\epsilon_e=0.15$, magnetic energy fraction $\epsilon_B=10^{-4}$, and electron energy distribution index $p=2.2$.}
\end{figure*}
\section{Conclusions and Discussion}
We have made temporal and spectral analyses of GRB 100418A. We show that the X-ray lightcurve is composed of two distinguished components. The first component ends with a steep decay segment, indicating that it is consistent with the internal origin. The late component is a smooth bump peaking at $\sim 10^5$ second post GRB trigger. This component is also detected in the optical bands. The possible physical origin of the second component has been discussed. We show that a late off-axis jet component with parameters, $E_{k,\rm iso}=10^{53}$ erg, $\theta_j=0.30$ rad, $n=0.05$ cm$^{-3}$, $\theta_{v}=0.315$ rad, $\epsilon_e=0.15$, $\epsilon_B=10^{-4}$, and $p=2.2$, can accurately fit the late hump of the afterglows.

No detection of any supernova associated with the nearby long GRB 060614 resulted in debate on the physical origin of this event, i.e., collapse of massive stars (Type II) or merger of compact stars (Type I) (e.g., Zhang 2006 and references therein). GRB 100418A is at a reshift of 0.624. Deep optical monitoring did not find any signature of SN light in the late optical lightcurve as seen for most GRBs at $z<1$ (see Zeh et al. 2004 for a full sample before the Swift era). It is possible that both GRBs 060614 and 100418A may be from a subclass of long GRBs without an accompanying SN. It is interesting that the temporal features of this event are similar to GRB 060614.

The circum burst environment is also critical to understanding the nature of a GRB (e.g., Xin et al. 2011). Two types of media are discussed in the literature, namely, an interstellar medium (ISM) with a constant density and a stellar wind with a density profile $n\propto r^{-2}$. A wind type of medium would undoubtedly indicate a massive star progenitor, since mergers of compact stars usually occur at the out-skirts of galaxies. A low-density medium is evidence for a burst from the merger of compact stars. We got $n=0.05$ cm$^{-3}$ in our model fit to the second bump\footnote{The medium density surrounding this event is reported as $n\geq 0.1$ cm$^{-3}$ by Marshall et al. 2011.}. This favors the scenario of a merger of compact stars as the progenitor of this event. However, with the classification method proposed by L\"{u} et al. (2010), GRB100418A is classified into the Type II group (log$\varepsilon=-0.57$) and GRB 060614 is marginally in the Type I group without considering its long extended emission. These results make it difficult to know the progenitors of these kinds of events. We should point out that one cannot confidently exclude the possibility that the two bursts may essentially have a different physical origin from both Type I and II, such as a stellar object being disrupted by a medium-mass black hole (Lu et al. 2008).

\normalem
\begin{acknowledgements}
We acknowledge the use of the public data from the Swift data archive. We appreciate valuable comments and suggestions from the referee. The work is supported by the National Natural Science Foundation of China (Grants Nos. 11025313, 10873002 and 11078008), National Basic Research Program of China (973 Program, Grant No. 2009CB824800), Chinese Academy of Sciences (Grant KJCXZ-YW-T19), Guangxi SHI-BAI-QIAN project (Grant No. 2007201), and Guangxi Science Foundation (2010GXNSFC013011, 2011-135).
\end{acknowledgements}

\label{lastpage}

\end{document}